\documentclass[twocolumn,aps,prl,multicol] {revtex4}
\usepackage{graphicx}
\begin{document}
\title{\begin{flushleft}Low-temperature,
 high-resolution magnetic force microscopy using a quartz tuning fork\end{flushleft}}
\author{\begin{flushleft}Yongho Seo, Paul Cadden-Zimansky and
 Venkat Chandrasekhar\vspace{-0.43cm}\end{flushleft}}
\affiliation{\begin{flushleft}\text{Department of Physics and
Astronomy, Northwestern University, Evanston, IL 60208,
USA}\vspace{-0.43cm}\end{flushleft}}
\date{\today}

\begin{abstract}
We have developed a low temperature, high resolution magnetic
force microscope (MFM) using a quartz tuning fork that can operate in a magnetic field. A tuning fork
with a spring constant of 1300 N/m mounted with a commercial MFM cantilever tip
was used. We have obtained high resolution images of the stray
magnetic fields exerted from grains with a spatial resolution of 15
nm and force resolution of 2 pN at 4.2 K. Tuning fork based
magnetic force microscopes have the potential to be used at
millikelvin temperatures due to their low power dissipation and
high force sensitivity.


\end{abstract}

\maketitle

\pagebreak

Since the possibility of a force microscope to detect magnetic
stray fields was first demonstrated by Martin and
Wickramashinghe \cite{Martin}, magnetic force microscopy (MFM)
has been developed as a high resolution magnetic imaging
tool \cite{Saenz}.  In a conventional MFM, optical detection is used to measure the vibrational amplitude and frequency of a cantilever. These micro-fabricated
cantilevers have small spring constants (0.1 $\sim$ 10 N/m)
leading to a high force sensitivity ($\sim$ 1 pN), comparable to the
magnetic force between a tip and sample. However, cantilever based MFM
has some weak points: (i)  inconvenient optical alignment, (ii) a tendency for the tip to crash into the surface due to the low stiffness of the cantilever, (iii) high resolution MFM is difficult because
the dithering amplitude is large ($\gtrsim$ 10 nm), and
(iv) the sample is exposed to laser light that may be detrimental for some applications.

One alternative method of optical detection that has been used for low temperature MFM is fiber
optic interferometry \cite{Moser,Allers}.   However, one of the disadvantages of the fiber-optic method is that
the design of the MFM scan head is necessarily complex as it requires more than two coarse approach
mechanisms in order to align the fiber optics and the
cantilever. To reduce the complexity of the MFM design, piezoresistive cantilever
detection has been used \cite{Yuan,Volodin}.  While piezoresistive cantilever detection has
advantages such as simple design and electrical detection without
optics, it has serious drawbacks such as low sensitivity, poor
spatial resolution and excessive heat dissipation.

Using a tuning
fork as a force transducer in a MFM overcomes many of the drawbacks of other designs: (i) Because the tuning fork sensor is stiff, the
tip mounted on the tuning fork is not as easily pulled to the sample surface by attractive forces. (ii) Since the spectral noise density of the tuning fork is $\rm
\sim 100 ~fm/ \sqrt{Hz}$ \cite{Giessibl}, the minimum dithering amplitude is much smaller than that of
a cantilever, allowing high
resolution imaging. (iii) As the tuning fork is a self-dithering and
self-sensing device, no optics are required and it is simple and
small. (iv) Because no light source is necessary and
the dissipated power can be reduced down to $\sim$ 1
pW \cite{Giessibl}, tuning fork based MFM can be operated in the dark and in low
temperature conditions.

Edwards {\it et al.} \cite{Edwards} first demonstrated a MFM using a tuning fork with a piece of cut Fe wire.
Todorovic and Schultz \cite{Todorovic} employed an etched nickel wire
as a tip in a similar geometry. They used a miniature tuning fork
having a stiffness $k$ of 2000 N/m and they obtained MFM images of
a hard disk. Both of these groups \cite{Edwards,Todorovic} achieved
reasonable spatial resolution.  However, the image contrast was not
satisfactory compared with that obtained by cantilever based MFM.
While the resonance frequency $f$ of a tuning fork is similar to that
of the cantilever, the quality factor $Q$ and the spring constant $k$ of a
tuning fork are typically 10$^2$ and 10$^4$ times larger than
those of the cantilever, respectively. Because the sensitivity of
MFM is proportional to $\sqrt{Qf/k}$ in terms of thermal noise,
the tuning fork based MFM is less sensitive than the cantilever
based MFM by a factor of 10 \cite{Todorovic}. This has been the most
serious limitation of the tuning fork based MFM.

In this Letter, we report on high resolution and high sensitivity MFM images
that were obtained using a tuning fork based MFM.  The increased resolution and sensitivity are due to the use of tuning forks with smaller spring constants, operation at low temperatures which increases the $Q$ of the tuning fork and improves the stability of the instrument, and to our technique of attaching commercial cantilever tips to the tuning fork, which minimizes the loading of the tuning fork, leading to a high $Q$ value and a concomitant increase in the sensitivity.

Each prong of the tuning fork we used is 2.2 mm
long, 190 $\mu$m thick and 100 $\mu$m wide. This
geometry of the tuning fork corresponds to a value of the spring
constant $k$ $\simeq$ 1300 N/m.  The technique for attaching the tip is similar to
that described by Rozhok \textit{et al.} \cite{Rozhok}.  A tip with height $15 \sim 20$
$\mu$m on a commercial MFM cantilever (Micromasch,
NSC36/Co-Cr) was used. The cantilever was cut
from its Si chip and glued at the end of a tuning fork prong with the tip facing
outwards using a home-made
micro-manipulator. The tip was magnetized by an electro-magnet.
After the tip was mounted on the tuning fork, the resonance
frequency decrease $\Delta f$ was about 20 Hz. According to
the mass loading effect of the quartz crystal microbalance
technique\cite{Stockbridge}, the added mass $\Delta m$ can be
estimated by $\Delta f/ f_0$ $\simeq \Delta m / m_0 $, where $m_0$
is the mass of the tuning fork and the original frequency of the
tuning fork $f_0$ is 32.768 kHz.  This gives a total added mass
including the epoxy, cantilever and tip of just 0.1 $\mu$g.

After the tip was mounted, the resonance full width at half maximum was
about 5 Hz  in air (1 Hz in vacuum). Its corresponding $Q$-value
was about 10$^4$ in air ($3\times 10^4$ in vacuum). These
$Q$-values were almost the same as those before the tip was
mounted. We attribute the remarkably small change in the $Q$-values to the very
small amount of glue used. Because the $Q$-value was roughly 100
times larger than that of a cantilever, it compensated for the high
stiffness of the tuning fork in terms of the sensitivity.

For the
coarse approach mechanism, we modified the walker designed by
Gupta and Ng \cite{Gupta} by replacing some materials in their
original design for operation at low temperatures. The housing material
(stainless steel in their design) was replaced by machinable ceramic (Macor) and
polished alumina plates were attached to the housing surface to minimize the friction between the
sapphire disks attached to the walker tube and the walker housing.  With other
choices of the contact materials (sapphire/glass or sapphire/brass), the walker
invariably froze on cooling even to liquid nitrogen temperatures.

The entire scan head was mounted on the end of an insert that could be cooled to
77 K and 4.2 K by dipping into liquid nitrogen and liquid helium respectively.
A magnetic field could also be applied by fitting the insert into a dewar equipped
with a two-axis magnet capable of fields of 3 T in the axial direction and 1 T in
the transverse direction.  The resonance frequency shift and phase shift were measured by
commercial phase-locked-loop electronics (easyPLL from Nanosurf). We
did not use any low temperature preamplifier. The tuning fork
signal was passed through a coaxial cable 1 m long and was fed
into a current-voltage amplifier located outside the insert.
The current-voltage amplifier consists of a 5 M$\Omega$ load
resistor and $\times$ 100 gain instrumentation amplifier. The
typical dithering amplitude was 5 nm (10 nm) at a drive voltage of
3.5 mV (10 mV) at 4.2 K (77K).

\begin{figure}
\includegraphics[width=8cm]{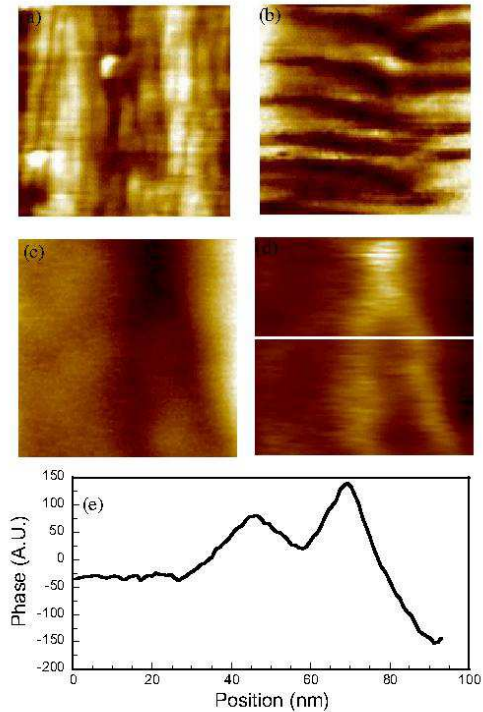}
 \caption{Representative topographic (a) and MFM (b) images obtained simultaneously by a tuning fork probe at 77 K.
The sample is a piece of a commercial hard disk. (c) and (d) represent higher resolution (94 nm $\times$ 94 nm) simultaneous
topographic (c) and MFM (d) images  of the same sample taken at 4.2 K.  (e) shows the line
profile across the white line shown in (d). }
\label{fig1}
\end{figure}

Figures 1(a) and 1(b) show representative topographic and MFM images obtained simultaneously by the
tuning fork probe at
77 K. The sample was a piece of a commercial hard disk in which the
magnetic recording layer is a cobalt film. The size of the images is 3
$\times$ 3 $\mu$m$^2$.  For
MFM scanning, constant height mode was employed with a lift height
50 nm. As there is little correlation between the two images, the MFM contrast
indeed comes from the magnetic stray field.

Since the dynamic MFM contrast represents the spatial second
derivative of the stray field or the force gradient, the frequency
shift in MFM corresponds to the gradient of the magnetic
force\cite{Dicarlo}
\begin{equation}
\Delta f \simeq {1 \over 2}{f \over k}{\partial F \over
\partial x},
\end{equation}
where $f$ is the resonance frequency (33 kHz), $k$ the spring
constant of the tuning fork (1300 N/m), and $F$ the magnetic force
between the tip and the sample. The experimental noise level of
the phase measurement was less than 0.5$^\circ$, which corresponds to a frequency
resolution of 1 mHz and a experimental force gradient resolution of
$10^{-4}$ N/m. Considering that the dithering amplitude of the
tuning fork was 10 nm, the experimental force resolution was 2 pN.

In order to evaluate the spatial resolution the same hard disk sample was
scanned over smaller areas with a 3.5 mV excitation at 4.2 K, corresponding to a dithering
amplitude of 5 nm.  Figures
1(c) and (d) show topographical and MFM images obtained
simultaneously over an area of 94 nm $\times$ 94 nm.
Lift mode scanning was applied for the MFM imaging by using the topographic profile to
raise the tip 20 nm over the device. Due to low thermal noise and drift, a very high
resolution
topographic image was obtained.  The maximum height
difference in Fig. 1(c) is 4 nm. The line profile between the
arrows on the MFM image is shown in Fig. 1(e).  Considering the narrowest width of peaks, the
spatial resolution of our microscope is about 15 nm. This
resolution is the same as the best resolution obtained with the
conventional cantilever based MFM\cite{Szmaja,Phillips} and much
better than that of piezoresistive cantilever based MFM
\cite{Yuan,Volodin}.

\begin{figure}
\includegraphics[width=8cm]{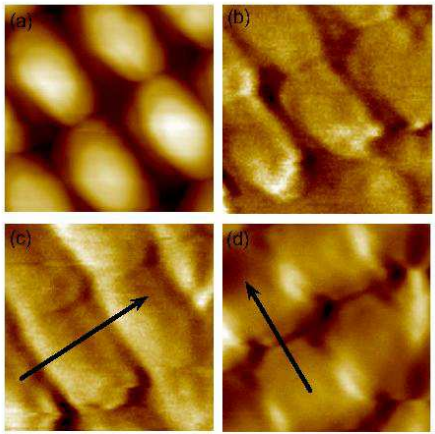}
\caption{1 $\mu$m $times$ 1 $\mu$m simultaneous topographic (a) and MFM (b) images of an array of elliptical permalloy particles at 4.2
K, with magnetic
field H = 0. (c), (d) MFM images of the same sample at 4.2 K with an applied field of 2.5 kG.  The magnetic field was applied in the direction shown by the arrows
in the figures.}
\end{figure}

In order to demonstrate MFM imaging in an external magnetic field, a permalloy (NiFe) sample consisting of an array of elliptical particles fabricated by e-beam lithography was imaged at 4.2 K with the external field applied in the plane of the permalloy film. Each elliptical particle has a long axis of 600 nm and short axis of 300 nm based on scanning electron micrograph images.  The topographic image of
the sample is shown in Fig. 2(a). The scanned area was 1
$\times$ 1 $\mu$m$^2$ and the excitation voltage was 7 mV. Figure
2(b) shows the MFM image of the same region without magnetic
field. Judging from the contrast in each particle, it appears that each
particle has its magnetization aligned along its long axis. Dramatic
changes of the MFM images occurred when a horizontal magnetic
field (H = 2.5 kG) was applied. This field is large enough to align the magnetization of
the sample as well as the tip.  Figures 2(c) and (d) show the resulting MFM images with the
direction of magnetic field denoted by the arrows.  Both images have stripe
patterns running at right angles to the magnetization. This is due to the attractive and
repulsive
interaction between the tip and sample that  depend on the relative location of the tip.

In conclusion, we have demonstrated a low temperature high
resolution MFM using a quartz tuning fork. We have obtained clear
images with a spatial resolution of 15 nm at 4.2 K. Tuning
fork base MFM is very promising candidate for millikelvin
temperature MFM, because of its low power dissipation, simple
design, and high spatial resolution.

This work was supported by the NSF through grant number ECS-0139936. We thank A. K. Gupta
and K. -W. Ng for helpful discussions about the low temperature coarse
approach mechanism.

\end{document}